\documentclass[preprints,article,accept,moreauthors,pdftex]{Definitions/mdpi}

\firstpage{1} 
\pubvolume{1}
\issuenum{1}
\articlenumber{0}
\pubyear{2023}
\copyrightyear{2023}
\datereceived{ } 
\daterevised{ } 
\dateaccepted{ } 
\datepublished{ } 
\hreflink{https://doi.org/} 

\Title{Pixel-to-Abundance Translation: Conditional
Generative Adversarial Networks Based on Patch Transformer for Hyperspectral Unmixing}

\TitleCitation{Title}

\Author{Li Wang$^{1}$\orcidA{}, Xiaohua Zhang $^{1,}$*, Longfei Li$^{1}$, Hongyun Meng $^{2}$ and Xianghai Cao $^{1}$}

\AuthorNames{Firstname Lastname, Firstname Lastname and Firstname Lastname}

\AuthorCitation{Lastname, F.; Lastname, F.; Lastname, F.}

\address{%
$^{1}$ \quad School of Artificial Intelligence, Xidian University; Key Laboratory of Intelligent Perception and Image Understanding of Ministry of Education, Xi'an 710071, China; liwang\_233@stu.xidian.edu.cn (L.W.);  20181214085@stu.xidian.edu.cn (F.L.); caoxh@xidian.edu.cn (H.C.);\\
$^{2}$ \quad School of Mathematics and Statistics, Xidian University, Xi'an 710071, China;  menghy@xidian.edu.cn (Y.M.)}

\corres{Correspondence: Xh\_zhang@mail.xidian.edu.cn (X.Z.)}

\abstract{Spectral unmixing is a significant challenge in hyperspectral image processing. Existing unmixing methods utilize prior knowledge about the abundance distribution to solve the regularization optimization problem, where the difficulty lies in choosing appropriate prior knowledge and solving the complex regularization optimization problem. To solve these problems, we propose a hyperspectral conditional generative adversarial network (HyperGAN) method as a generic unmixing framework, based on the following assumption: the unmixing process from pixel to abundance can be regarded as a transformation of two modalities with an internal specific relationship. The proposed HyperGAN is composed of a generator and discriminator, the former completes the modal conversion from mixed hyperspectral pixel patch to the abundance of corresponding endmember of the central pixel and the latter is used to distinguish whether the distribution and structure of generated abundance are the same as the true ones. We propose hyperspectral image (HSI) Patch Transformer as the main component of the generator, which utilize adaptive attention score to capture the internal pixels correlation of the HSI patch and leverage the spatial-spectral information in a fine-grained way to achieve optimization of the unmixing process. Experiments on synthetic data and real hyperspectral data achieve impressive results compared to state-of-the-art competitors.}


\begin{document}




\section{Introduction}
Hyperspectral image (HSI) has been widely used in mineral detection, environmental disaster monitoring, military, and many other fields \cite{ref1, ref3, ref4, ref5}, and the research works on it have attracted considerable attention. HSI is obtained by imaging spectrometers, having hundreds of narrow band spectral information in each pixel, and the abundant spectral information significantly enhances its ability to detect material attribute information. Nevertheless, as the low spatial resolution of HSI, each pixel typically is a mixture of several pure substances. These mixed observations hinder the application and development of HSI analysis technology. Therefore, hyperspectral unmixing (HSU), which aims to estimate the pure substances (endmembers) of the mixed pixels in HSI as well as their fractional abundances, is a significant and challenging task in the area of hyperspectral remote sensing.

For a given instantaneous field of view (IFOV), differences in surface feature type, spatial distribution structure, lighting conditions, and application requirements will result in various hyperspectral mixing methods, and demand corresponding unmixing models. Among the unmixing models, the linear mixture model (LMM) assumes that the features on the reflective surface are distributed in a tessellated pattern, and the incident light is directly received by the remote sensing instrument after interacting with one kind of material at a time. LMM is most broadly used in HSU for its algorithmic simplicity and physical significance. Nonnegative matrix factorization (NMF) achieves spectral unmixing by decomposing HSI into the matrix of endmembers and the abundance matrix \cite{ref7, ref8, ref9, ref10, ref11, ref12}. Sparse unmixing by variable splitting and augmented Lagrangian (SUnSAL) is a classical sparse unmixing algorithm \cite{ref13} that searches the most sparse linear endmember combination of HSI from the known complete endmember spectral library. Bioucas \emph{et al}. \cite{ref14} take the lead in introducing total variation regularization into sparse unmixing and propose a sparse unmixing algorithm (SUnSAL-TV) based on total variation space regularization. It promotes spatial correlation through the first-order neighborhood information of pixels and integrates spatial information into the sparse unmixing model to promote the smoothness of homogeneous regions of images, which brings a new concept for sparse unmixing based on spatial information.

However, in the complex case of some special ground objects distribution, the light has been repeatedly scattered between substances at different heights, so the nonlinear mixture model has been developed according to the actual situation to model multiple interactions of the light. Bilinear mixture model (BMM) considers the secondary scattering between substances, but ignores the higher-order scattering of more than three times, and expresses the nonlinear scattering effect by the Hadamard product and nonlinear coefficient between endmember vectors. The bilinear mixing models like the generalized bilinear model \cite{ref15} describes the interaction of an incident ray of light with two substances. Another bilinear model methodology is presented in \cite{ref16}, which proposes a nonlinear transformation of the spectrum generated by the LMM to introduce nonlinearity. The transformation proposed is a polynomial of degree two, leading to the polynomial post-nonlinear model (PPNM). For special scenes such as complex urban buildings where multiple scattering is the dominant component, higher-order scattering other than the second scattering between the endmembers in the BMM should be considered. Heylen \emph{et al}. \cite{ref17} introduce a parameter $P$ to describe the probability of high-order interaction between light rays and substances based on PPNM, thus extending PPNM to the multilinear mixing model (MLM). The radiative transfer (RT)-based Hapke model \cite{ref18} can accurately describe the light scattering from matter in the observed scene and define complex nonlinear mixing mechanisms for the spectrum, but it makes the unmixing problem intractable. To bypass the difficulty of defining the spectral mixing mechanism in the original space where the data are located, the kernel method\cite{ref64} in machine learning is used to map the hyperspectral pixel from the original space to the high-dimensional feature space via some kernel functions. In the feature space, the original nonlinear problem can be solved using linear algorithms. Chen et al. \cite{ref20} propose a kernel-based hyperspectral mixing model (K-Hype), which adds a nonlinear perturbation term to the linear mixing model, where the perturbation term depends only on the endmember. Multiple-kernel learning-based model (MKHS) \cite{ref56} integrates a reproducing kernel Hilbert space (RKHS) spanned by different base kernels, providing more capability in dealing with nonlinear unmixing problems than the traditional RKHS single kernel learning. It describes the higher-order interactions between endmembers. Although the kernel-based algorithms provide flexible nonlinear modeling, the selection of kernel functions and related parameters largely determine the effectiveness of the model, which limits the application of these methods. The above mentioned are unmixing methods based on regularization, which require setting the corresponding prior knowledge and regularization terms for hyperspectral data under different scenarios and need to solve new optimization problems each time to unmix, thus parametrically complex.

Most of the aforementioned models need to set prior parameters according to specific scenarios. The data-driven deep learning methods have been proposed to address the unmixing problem recently, which avoids complex physical models to some extent. For instance, the autoencoder network (AE) is widely applied to solve the unmixing problem. AE is a network that requires minimization of the difference between input original hyperspectral pixels and output reconstruction hyperspectral pixels. AE includes an encoder and a decoder, in which the encoder dimensionally compresses the mixed hyperspectral pixels to obtain features, and feeds the features to the decoder for final reconstruction into the original mixed hyperspectral pixels, obtaining the endmembers and the corresponding abundances \cite{ref21, ref22, ref23}. In such an unmixing process, AE still needs to consider a particular mixing model when setting the loss function, such as the linear mixing model \cite{ref24}, nonlinear mixing model \cite{ref25}, bilinear mixing model \cite{ref26}, etc. However, for the mixed spectrum formed by the higher-order interaction of incident radiation between endmembers, its optical properties rely on lots of parameters, for instance, the quantity and proportion of material components, the size, shape, direction, and distribution of particles, absorption, and scattering characteristics. It is difficult for AE to obtain the corresponding prior knowledge and accurately model the mixing characteristics of this process \cite{ref27}. Therefore, AE has a certain limitation of unmixing scenes and the parameter setting problem of the model still exists. To overcome these drawbacks, we adopt the generative adversarial network (GAN) \cite{ref28} to achieve hyperspectral unmixing. In this paper, we propose conditional generative adversarial networks based on HSI Patch Transformer (HyperGAN) for hyperspectral unmixing, where the discriminator networks can automatically learn weight parameters, to replace a manually set loss function for different complex mixing scenarios. The proposed method uses an adversarial strategy to discriminate the true and false output abundance by a discriminator, and the generator aims to generate the abundance that can deceive the discriminator. This process is equivalent to the generator automatically simulating a suitable unmixing model to generate an abundance close to the true abundance. Also, this abundance generation process can be considered as an image-to-image translation process. For many image-related problems in computer vision, GAN is more like a "translating" method that translates the given input image to its corresponding output \cite{ref29, ref30, ref31}. Therefore, we define the hyperspectral unmixing task as translating the pixel representation of HSI into an abundance representation using a limited number of abundance labels, namely, the pixel-to-abundance translation task.

In addition, the convolutional neural network (CNN) is the most widely utilized method to catch spatial information from HSI to assist HSU \cite{ref32, ref33, ref34}. By operating the convolution kernel on the HSI, the features of the target pixel and its surrounding pixels are equally extracted to serve for target pixel unmixing, which ignores the magnitude of the correlation between the target pixel and surrounding pixels. To utilize the hyperspectral spatial information more precisely, we propose the HSI Patch Transformer for abundance generation. The transformer is a type of neural network mainly consisting of the self-attention layer \cite{ref41} that can provide the relationships between different features. Transformer is commonly applied in natural language processing (NLP) tasks, e.g., the famous BERT \cite{ref35} and GPT-3 \cite{ref36} models. The power of these transformer models inspires many researchers to investigate the use of transformer for visual tasks. Different from the data in NLP tasks, there exists a semantic gap between input images and the ground-truth labels in computer vision (CV) tasks. To this end, Dosovitskiy \emph{et al}. \cite{ref37} develop the ViT, which paves the way for transferring the success of transformer-based NLP models for CV tasks. In this article, the HSI Patch Transformer is presented, which divides the HSI into HSI patches and each pixel in the patch is a visual sequence. Then, the attention score can be naturally calculated between the central pixel and any other pixels of the patch for generating effective feature representations of the central pixel unmixing task. Compared with CNN, self-attention finds relevant pixels through adaptive attention score. The range and shape of the receptive field are no longer manually delineated but learned automatically by the HSI Patch Transformer. It selects feature threshold from candidate feature threshold for high-dimensional feature crossover, in which the selected feature threshold has a strong correlation with the current feature threshold. In this way, for the unmixing task of the central pixel, effective spatial information and interfering spatial information can be distinguished.

\begin{figure}[H]
\begin{adjustwidth}{-\extralength}{0cm}
\centering
\includegraphics[width=15cm]{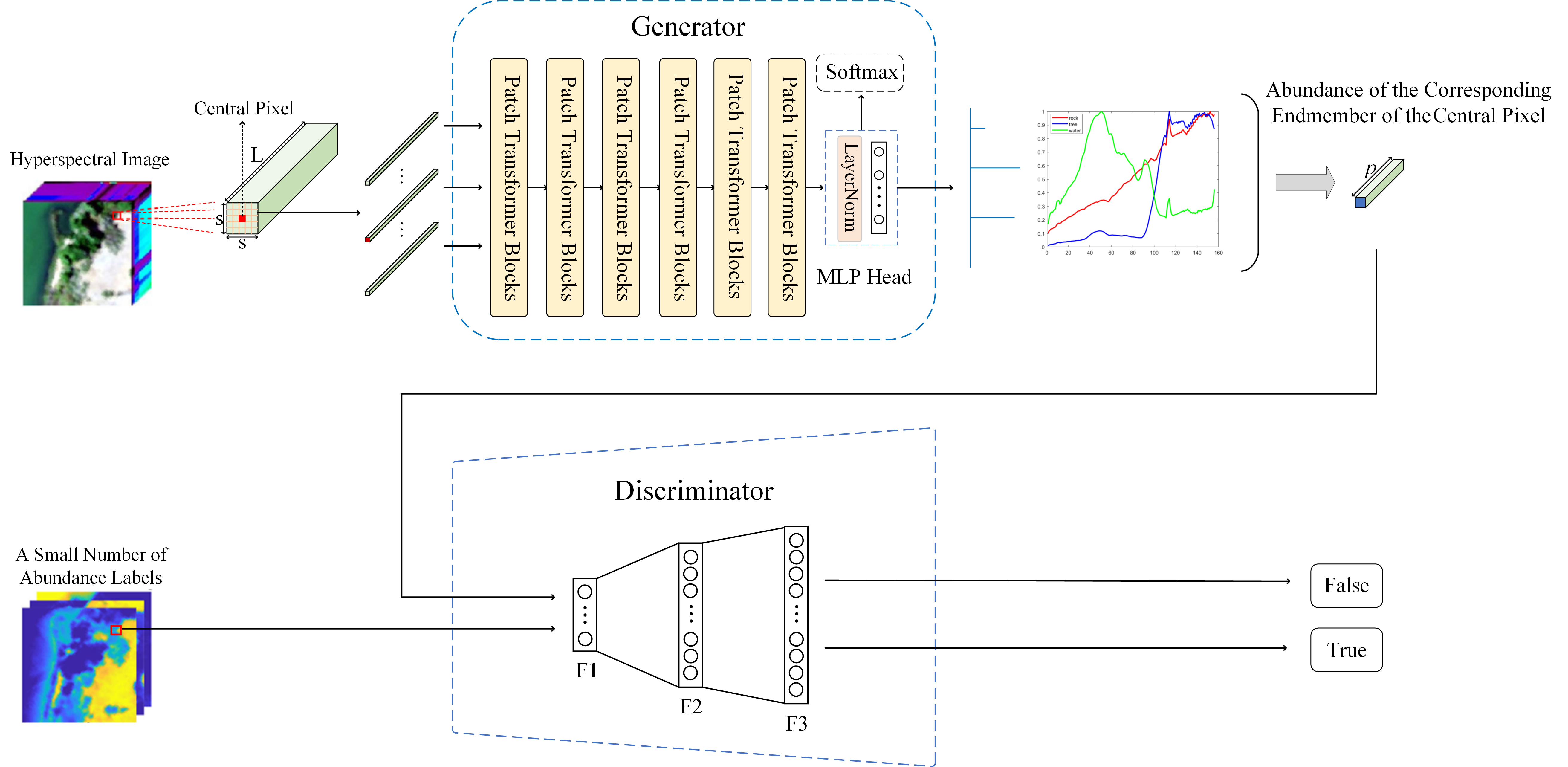}
\end{adjustwidth}
\caption{A schematic of the proposed HyperGAN method.\label{Fig. 1}}
\end{figure}

Lastly, as shown in Fig. \ref{Fig. 1}, we incorporate the HSI Patch Transformer into GAN, to develop the conditional generative adversarial networks  for a generic framework for the unmixing rather than rely on prior mixing scenario. This proposed method uses the generator to approximate the mapping from HSI pixels to the abundance of corresponding endmember of the central hyperspectral pixel to achieve unmixing, instead of needing to solve optimization problems as in the case of regularization unmixing. A discriminator is used to replace the complex regularization constraints to discriminate  whether the generated abundance is true or not. The major contributions are concluded as:
\begin{itemize}
\item[1)]
The proposed HyperGAN method provides a new perspective for HSU, and the process of obtaining estimated abundance from hyperspectral pixel can be viewed as modal transformation, where the abundance is a manifold representation in the low-dimensional space of the hyperspectral pixel. The discriminator and generator of HyperGAN play against each other through an adversarial strategy. The generator has to generate the abundance of corresponding endmember to deceive the discriminator, and the discriminator network has to discriminate the true and false abundance. In this unmixing process, the discriminator network replaces the loss function requiring sophisticated prior regularization terms, and the generator can automatically learn the appropriate unmixing model to generate the abundances. It eliminates the burden of introducing regularization terms and prior knowledge for different HSU scenarios and avoid solving complex regularization optimization problem.
\item[2)]
We propose the HSI Patch Transformer to incorporate spatial information in unmixing, by inputting HSI patches to generate the abundance of its central hyperspectral pixel. The weight of fusion features in the HSI Patch Transformer is dynamic and contains the calculation results of correlation between domain pixels and central pixel, which depends to some extent on the characteristics of the HSI data itself and is thus adaptive. Therefore, HSI Patch Transformer utilizes spatial-spectral information of hyperspectral pixels in a finer granularity way.
\item[3)]
Hyperspectral image data are not composed of isolated hyperspectral pixels, but of pixels with certain spatial correlation. We propose a data synthesis method based on superpixel segmentation and random split to get synthetic data with a spatial structure to further evaluate the effectiveness of our model for unmixing. The common hyperspectral image data synthesis method synthesizes individual pixels without giving consideration to the spatial structure of the HSI. In contrast, our proposed hyperspectral data synthesis method is more realistic.
\end{itemize}

The remaining sections of our article are structured as follows. Section II is the problem formulation about the transformation of pixel-to-abundance. The proposed HyperGAN unmixing methodology and hyperspectral data synthesis methodology are introduced in Section III. The experiments analysis as well as the results on real HSI data and synthetic data are given in Section IV. Finally, conclusions are stated in Section V.

\section{Unmixing Mechanism Based on Modality Transformation}

This section defines the proposed mixture model and notation. Normal font \emph{x}  and \emph{X} denote scalars. Boldface small letters $\boldsymbol{x}$ denotes vectors. All vectors are column vectors. Boldface capital letters $\boldsymbol{X}$ denote matrices. Let $\boldsymbol{X}=[\boldsymbol{x}_1,\boldsymbol{x}_2,\cdots,{\boldsymbol{x}}_{\emph{N}}]\in{\mathbb{R}^{L\times{N}}}$ be the HSI possessing $N={r}\times{c}$ pixels, and the \emph{i}th pixel (${\boldsymbol{x}}_{\emph{i}}=[\emph{x}_{\emph{i},1},\emph{x}_{\emph{i},2},\cdots,\emph{x}_{\emph{i},\emph{L}}]^T\in\mathbb{R}^{L\times{1}}$) contains $L$ spectral bands, where r and c are the number of rows and columns of the HSI respectively and $\emph{i}=1,\cdots,\emph{N}$;
$\boldsymbol{M}=[\boldsymbol{m}_1,\boldsymbol{m}_2,\cdots,{\boldsymbol{m}}_{\emph{p}}]\in{\mathbb{R}^{L\times{p}}}$ represents the endmember matrix with each column denoting one of the $p$ endmembers, and $\boldsymbol{A}=[\boldsymbol{a}_1,\boldsymbol{a}_2,\cdots,{\boldsymbol{a}}_{\emph{p}}]\in{\mathbb{R}^{p\times{N}}}$ is the corresponding abundance map with the \emph{i}th column indicating the abundance vector for \emph{i}th pixel (${\boldsymbol{a}}_{\emph{i}}=[\emph{a}_{\emph{i},1},\emph{a}_{\emph{i},2},\cdots,\emph{a}_{\emph{i},\emph{p}}]^T\in\mathbb{R}^{p\times{1}}$). Considering the general mixing mechanism \cite{ref38}, the \emph{i}th pixel can be expressed as
\begin{equation}
\boldsymbol{x}_\emph{i}=\Phi(\boldsymbol{M},\boldsymbol{a}_\emph{i})+\boldsymbol{n}_\emph{i}
\end{equation}
where ${\Phi}$ represents the hidden function, defining the sophisticated relationship between the endmembers of the matrix $\boldsymbol{M}$ parameterized by abundance $\boldsymbol{a}_\emph{i}$, and $\boldsymbol{n}_\emph{i}$  is the natural noise of observed single pixel $\boldsymbol{x}_i$. With regard to HSI, the general mixing mechanism represented by (1) can be extended as
\begin{equation}
\boldsymbol{X}=\Phi(\boldsymbol{M},\boldsymbol{A})+\boldsymbol{N}
\end{equation}
where the abundance map is required to satisfy two physical constraints, that is: the abundance non-negativity constraint (ANC), i.e., $\emph{a}_{\emph{i},\emph{j}}\ge{0}$, and the abundance sum-to-one constraint (ASC), i.e., $\sum_{j=1}^{p}a_{i,j}=1$. ANC proves that all elements of the abundance map must be non-negative, while ASC requires that the sum of the abundance for each pixel is equal to one. When we move the additive noise term to the left side of (2), the $\boldsymbol{X}+(-\boldsymbol{N})$ term is denoted as $\boldsymbol{X}_{noise}$. Consequently, our model the unmixing problem as
\begin{equation}
\boldsymbol{X}_{noise}=\Phi(\boldsymbol{M},\boldsymbol{A})
\end{equation}
where $\boldsymbol{X}_{noise}\in{\mathbb{R}}^{L\times{N}}$ , and $\boldsymbol{A}\in{\mathbb{R}}^{p\times{N}}$.

\begin{figure}[h]
\centering
\includegraphics[width=2.5in]{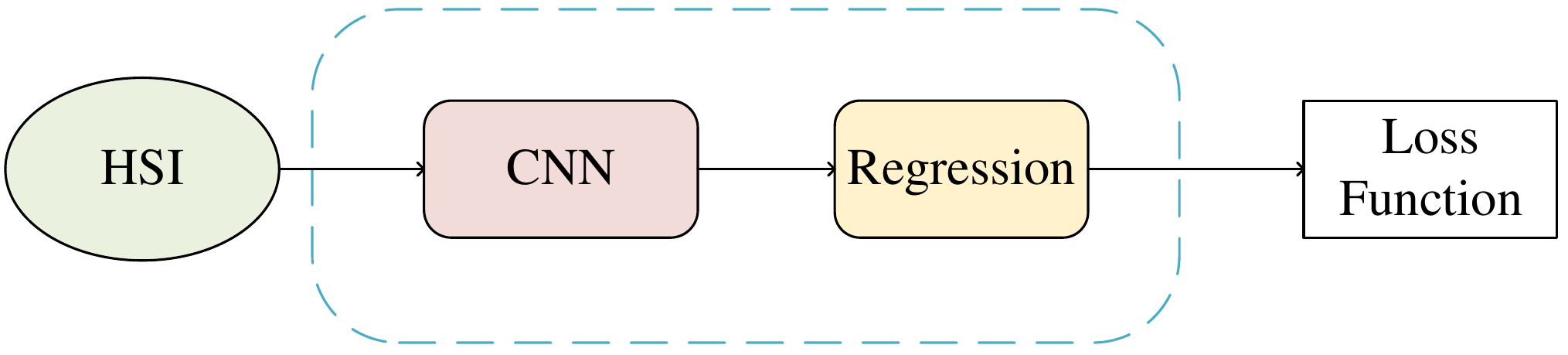}
\caption{Common hyperspectral unmixing idea framework.}
\label{Fig. 2}
\end{figure}
\begin{figure}[h]
\centering
\includegraphics[width=2.77in]{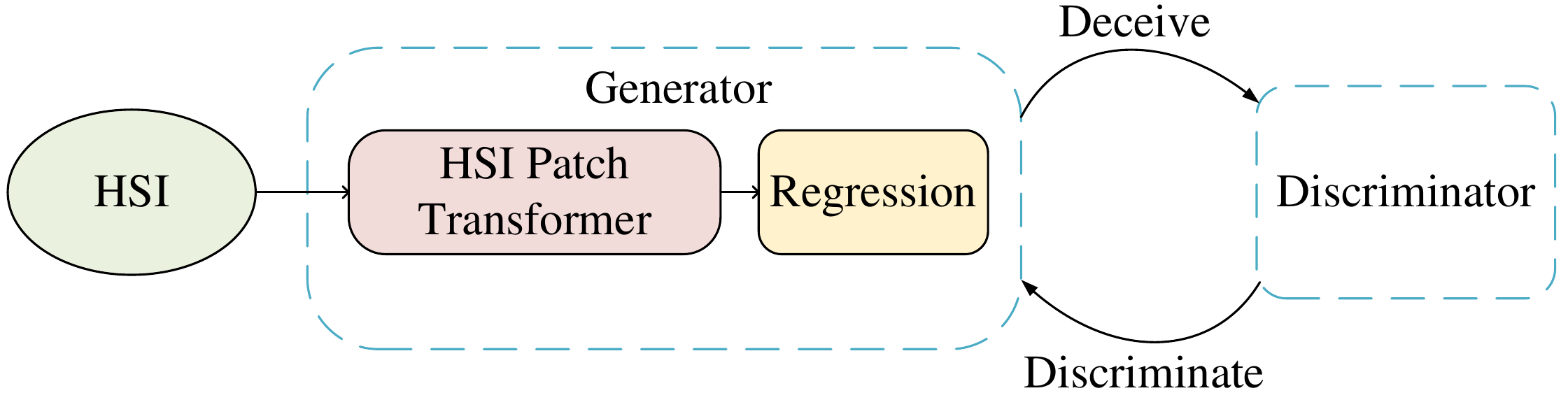}
\caption{The hyperspectral unmixing idea framework of the proposed method.}
\label{Fig. 3}
\end{figure}

For an mixing problem like (3), the framework of most deep learning unmixing methods is shown in Fig. \ref{Fig. 2}, where a deep learning network extracts features from HSI and regresses to obtain the abundance. This process learns the mapping relations between $\boldsymbol{X}_{noise}$ and the corresponding $\boldsymbol{A}$. In this case, the corresponding regularization loss function is required for training the unmixing network. For instance, the orthogonal sparse prior-based AE \cite{ref49} using Laplacian loss. Nevertheless, a specific prior loss function cannot satisfy different hyperspectral data unmixing scenarios. In addition, the extraction of effective features is a significantly critical step. To reduce noise and utilize the structural similarity of the abundance vector of adjacent pixels \cite{ref65}, most approaches use CNN to extract spatial spectral information features of HSI patches. Let $\boldsymbol{X}_{patch}=[\boldsymbol{x}_1,\boldsymbol{x}_2,\cdots,{\boldsymbol{x}}_{\emph{k}}]$ be the HSI patch, where \emph{k} is the size of the HSI patch. The CNN features can be approximated as 
\begin{equation}
\sigma(w_1\boldsymbol{x}_1+w_2\boldsymbol{x}_2+,\cdots,+w_{cen}\boldsymbol{x}_{cen}+,\cdots,+w_k\boldsymbol{x}_k)=y_i
\end{equation}
where $\sigma(x)$ denotes the activation function applied in the network and $cen$ means central, while $w_i$ and $y_i$ are the fused weights and the output feature of the CNN, respectively. After the network is well trained, the weights $w_i$ are fixed when the CNN extracts the features of different HSI patches. However, the structure of HSI patches is multivariate \cite{ref65} and the similarity between the domain hyperspectral pixels and central hyperspectral pixel in different HSI patches is different, which means that the weights in the above equation should be adaptive.

Then, as denoted in (3), $\boldsymbol{A}$ can be considered as a representation of $\boldsymbol{X}_{noise}$ in the new abundance space. And the column vectors of matrix $\boldsymbol{M}$, i.e., the endmembers, form a set of basis vectors of the abundance space. That is, $\boldsymbol{X}_{noise}$ and the corresponding abundance matrix $\boldsymbol{A}$ are both the representation of the HSI, in which the unmixing and mixing process can be regarded as a transformation process of two modalities.

Therefore, in this article, we construct an unmixing framework based on HSI Patch Transformer conditional GAN shown in Fig. \ref{Fig. 1}. It is a generic framework for the unmixing of different HSI, which regards the unmixing problem as a modality transformation problem. In this framework, the discriminator network that automatically learns parameters according to different unmixing scenarios replaces the sophisticated loss function which requires prior knowledge as regularization constraints. The mutual rivalry between the generator and discriminator makes the generator automatically simulate the appropriate unmixing pattern for different hyperspectral data and obtain the abundance of the corresponding endmember. The discriminator is used to discriminate whether the true abundance and the generated abundance follow the same distribution.

For the feature extraction step, we use HSI Patch Transformer, which has an adaptive attention influence factor when fusing features, i.e., the contribution of features is determined by  calculating the correlation  between the HSI patch domain pixels and the target central pixel, to obtain the spatial distribution information of the HSI patch abundance and improve the central pixel unmixing performance. Compared to CNN, it can be modeled as
\begin{equation}
\sigma(c_1\boldsymbol{x}_1+c_2\boldsymbol{x}_2+,\cdots,+c_{cen}\boldsymbol{x}_{cen}+,\cdots,+c_k\boldsymbol{x}_k)=y_i
\end{equation}
where $c_i$ is the weight of the HSI Patch Transformer. Unlike CNN, the weights are dynamic and contain the calculation results of the correlation between domain pixels and central pixel. The value of the weights depends to some extent on the characteristics of the HSI data itself and is adaptive. therefore, HSI Patch Transformer utilizes spatial spectral information of hyperspectral pixels at a finer granularity. 
\begin{figure*}[!t]
\centering
\includegraphics[width=5in]{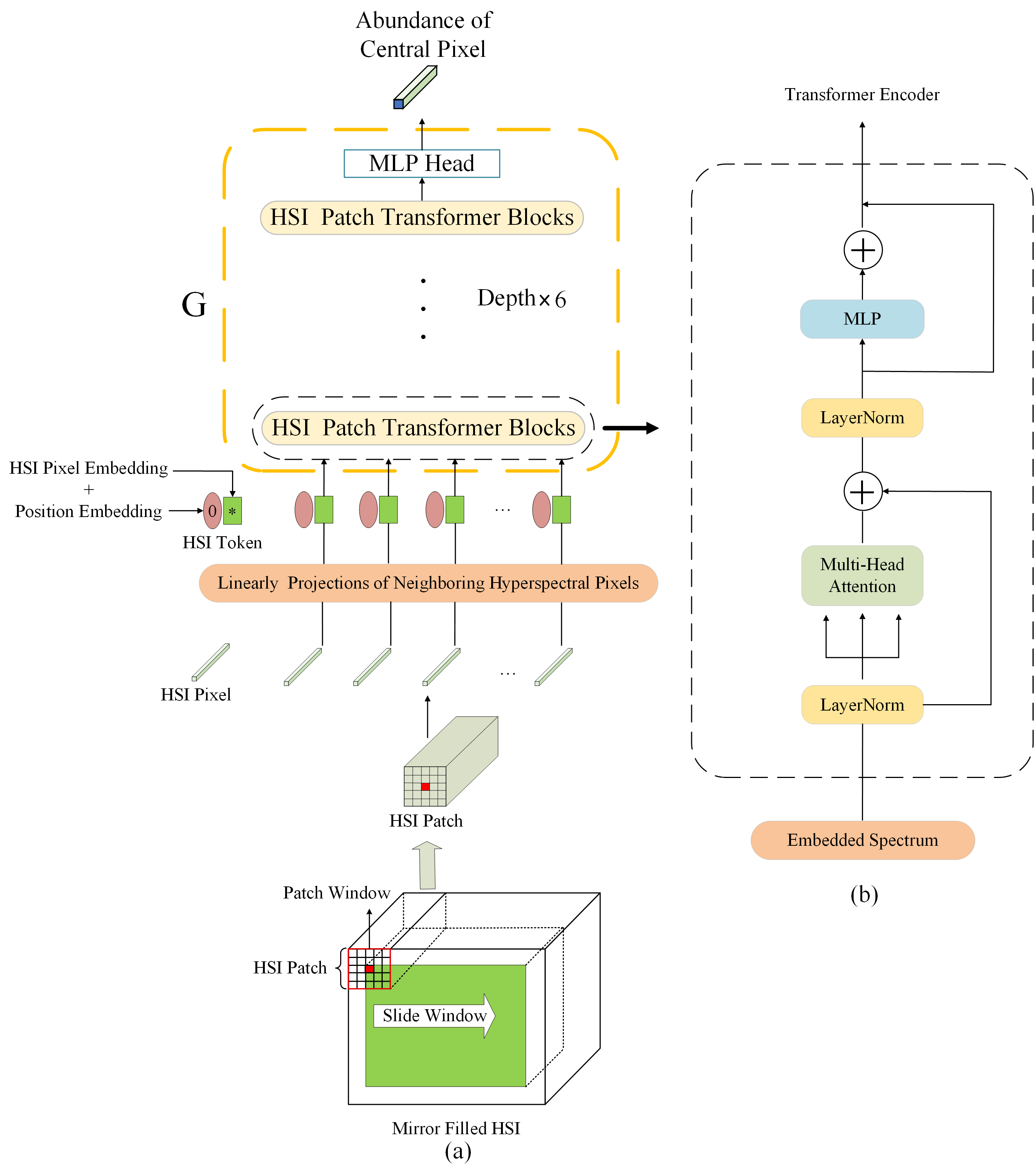}
\caption{(a) The HSI patch obtained by sliding window operating on the mirror-filled HSI. (b) The structure of HSI Patch Transformer blocks.}
\label{Fig. 4}
\end{figure*}

\section{The Proposed Method}
In this section, we present a comprehensive description of the proposed method. The conditional generation adversarial network based on the HSI Patch Transformer is proposed to solve the hyperspectral unmixing problem, and the data synthesis method based on superpixel segmentation and random split is proposed to synthesize hyperspectral data.
\subsection{Hyperspectral Unmixing Structure Based on Generative Adversarial Learning}
As shown in Fig. \ref{Fig. 1}, the proposed HyperGAN includes two parts, i.e., generator (G) and discriminator (D). G is composed of six HSI Patch Transformer blocks and a multilayer perceptron (MLP) head, while D is composed of three fully connected layers. In this algorithm, G takes a random image patch from HSI as input, and its output needs to emulate as much as possible the true abundance of the corresponding endmember of the central pixel in the corresponding HSI patch. The input of D is the true abundance or the generated abundance of the G output, and the purpose of D is to distinguish the generated abundance of the G output from the true abundance as much as possible.  The G, on the other hand, tries to deceive D as much as possible. The two networks confront each other and continuously adjust the parameters, with the ultimate goal of making the discriminator network unable to determine whether the generative network output is true or not. That means, our model achieves mapping as $\emph{G}:\boldsymbol{X}_{noise}\in{\mathbb{R}}^{L\times{N}}\to\boldsymbol{A}\in{\mathbb{R}}^{p\times{N}}$, which completes the supervised unmixing. In addition, for the HSI patches with natural random noise, the noise increases the sample diversity of HSI patches, and the pattern collapse problem of HyperGAN \cite{ref39} can be mitigated when it is used as conditional input, which allows HyperGAN to generate high-quality unmixing abundance. Due to the manipulation of HSI image patches rather than individual hyperspectral pixel, our method is able to exploit the spatial information in HSI. Finally, as for the Softmax functional layers, the Softmax layer of the MLP head is designed to meet the physical constraints of the generated abundance, i.e., the ANC and ASC, respectively.

\noindent  {\bf{Objective.}} 
Traditional GANs assume the discriminator as a classifier with a sigmoid cross-entropy loss function \cite{ref28}. Nevertheless, Mao \emph{et al}. \cite{ref40} found that the cross-entropy loss function may cause the problem of gradient disappearance during network training. Therefore, we adopt the least square loss function \cite{ref40} in our method. 1 and 0 are designed labels for the true abundance and generated abundance, respectively. This loss function improves two problems of GAN, namely, the traditional GAN generator is not effective enough, and the training process is very unstable \cite{ref39}. Moreover, to make the generated abundance close to the real abundance and guide the generator to generate the abundance, $L_1$ regularization is added and the abundance label is introduced as the constraint term. Thus, the full objective is:
\begin{equation}
\begin{aligned}
\min _{D} V_{HyperG A N}(D)&=\frac{1}{2} E_{\boldsymbol{a}\sim p_{\text {data }}(\boldsymbol{a})}\left[(D(\boldsymbol{a})-1)^{2}\right]\\
&+\frac{1}{2} E_{\boldsymbol{x} \sim p_{\text {data }}(\boldsymbol{x})}\left[D^{2}(G(\boldsymbol{x}))\right] \\
\min _{G} V_{HyperG A N}(G)&=E_{\boldsymbol{x} \sim p_{\text {data}}(\boldsymbol{x})}\left[(D(G(x))-1)^{2}\right]\\
&+\lambda_{\text {cor}} E_{\boldsymbol{x} \sim p_{\text {data }(\boldsymbol{x})}}\left[\left\|G(\boldsymbol{x})-\boldsymbol{a}^{*}\right\|_{1}\right]
\end{aligned}
\end{equation}
where $\boldsymbol{x}$ is the HSI patch with natural noise, $\boldsymbol{a}$ is the corresponding true abundance of the central HSI pixel in the patch, $G(\boldsymbol{x})$ is the generated abundance, $\boldsymbol{a}^{*}$ is the true abundance label, $\lambda_{\text {cor}}$ is a hyperparameter that acts as a correction coefficient for generator to guide the process of generating abundance.
\begin{figure}[!t]
\centering
\includegraphics[width=8.5 cm]{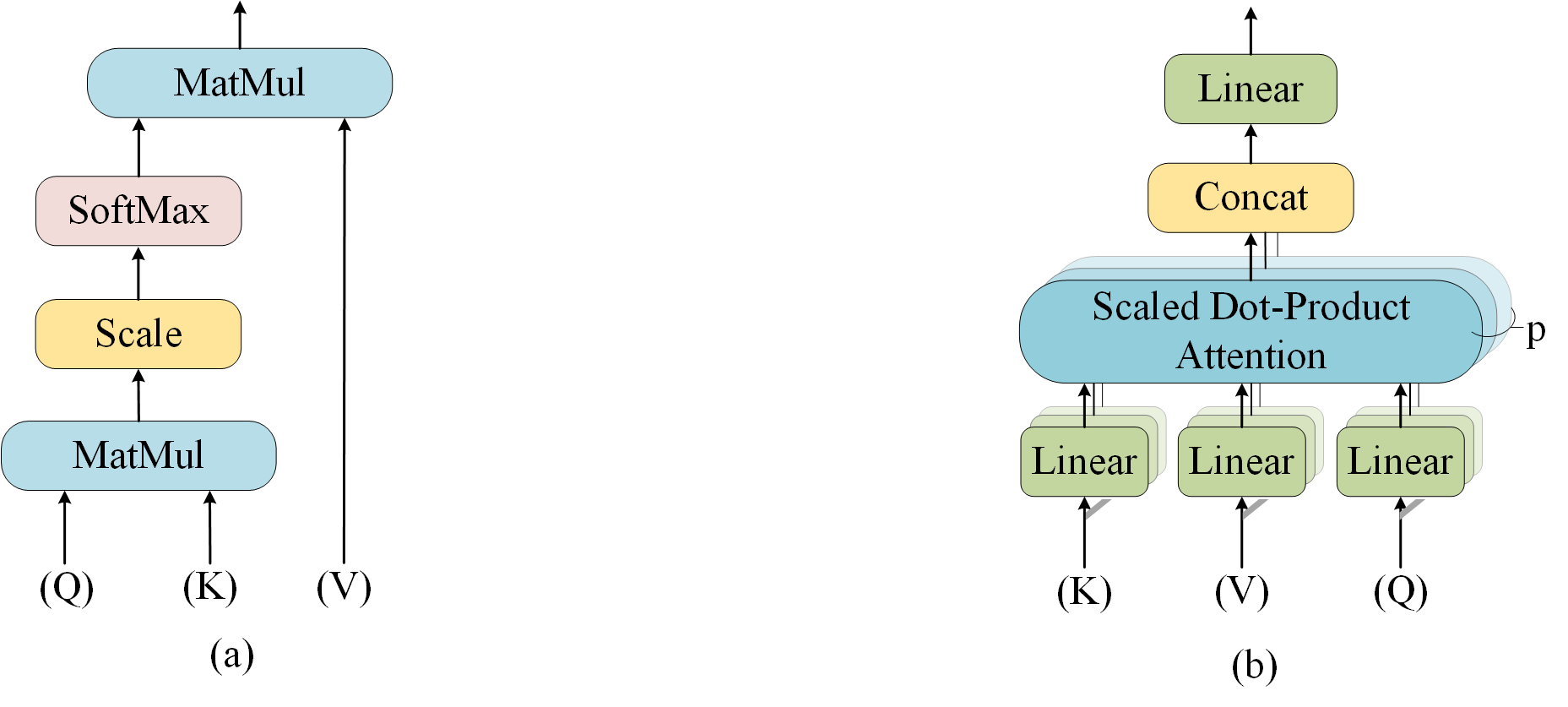}
\caption{(a) Self Attention. (b) Multi-Head Attention.}
\label{Fig. 5}
\end{figure}

\subsection{HSI Patch Transformer Blocks}
In this work, we develop a HSI Patch Transformer blocks based generator that utilizes transformer blocks to incorporate spectral-spatial information, especially the abundant channel information of HSI pixels, and achieves better unmixing performance. The specific structure of HSI Patch Transformer blocks is present in Fig. \ref{Fig. 4}(b), which mainly contains the HSI patch input and multi-head attention. The skip connection is vital in the HSI transformer blocks. It may be well interpreted through the use of "residuals" to allow better propagation of gradients, or enhanced "memory" to decrease forgetting or loss of critical information \cite{ref55}.

\noindent  {\bf{HSI Patch Input.}}
HSI patch is obtained by operating a $5\times5$ sliding window on the HSI with a step size of 1, which is illustrated in Fig. \ref{Fig. 4}(a). The $5\times5$ window size is set by experimental analysis. Every pixel in the HSI is expected to be the center of a patch. As for edge pixels, mirror padding of the HSI to get the appropriate size is necessary. Apart from this, to go through all pixels, the slide step of the window is set to 1. HSI patch input allows the model to take full advantage of HSI spatial-spectral information.

\noindent {\bf{Multi-Head Attention.}}
HSI Patch Transformer blocks rely heavily on the use of multi-head attention \cite{ref41}, where multi-head attention is a deformation of self-attention \cite{ref57}. In practice, let the HSI image patch be $\boldsymbol{X}=[\boldsymbol{x}_1,\boldsymbol{x}_2,\cdots,{\boldsymbol{x}}_{\emph{k}}]\in{\mathbb{R}^{(L\times{s}\times{s})}}$, where $s\times{s}$ is the HSI spatial patch size and $L$ is the number of spectral bands. The hyperspectral pixels in the HSI patch are linearly projected to obtain the corresponding HSI pixel embedding and then add the randomly generated position embedding showing the pixel position discrepancy to obtain the HSI token as the input $\boldsymbol{X}_{token}$ to the HSI Patch Transformer blocks. We conduct the linear transformation on $\boldsymbol{X}_{token}$, as shown in the following equations:
\begin{equation}
    \begin{split}
    \boldsymbol{Q}&=\boldsymbol{W}_{\boldsymbol{Q}}\boldsymbol{X}_{token}\\
    \boldsymbol{K}&=\boldsymbol{W}_{\boldsymbol{K}}\boldsymbol{X}_{token}\\
    \boldsymbol{V}&=\boldsymbol{W}_{\boldsymbol{V}}\boldsymbol{X}_{token}
    \end{split}
\end{equation}
where $\boldsymbol{W}_Q$, $\boldsymbol{W}_K$, and $\boldsymbol{W}_V$ are the weight matrices of the network that need to be learned to obtain, and the query vector sequence ($\boldsymbol{Q}$), the key vector sequence ($\boldsymbol{K}$), and the value vector sequence ($\boldsymbol{V}$) are obtained by this $X_{token}$ transformation. The self-attention is performed as Fig. \ref{Fig. 5}(a), which includes inner product and scaling operations. The input consists of $\boldsymbol{Q}$, $\boldsymbol{K}$, and $\boldsymbol{V}$ of dimension $d$. $\boldsymbol{Q}$ is used to match other $\boldsymbol{K}$, $\boldsymbol{K}$ is to be matched, and $\boldsymbol{V}$ is the information of $\boldsymbol{X}$ to be extracted. The self-attention outputs are caculated as:
\begin{equation}
\emph{Self-attention}(\boldsymbol{Q},\boldsymbol{K},\boldsymbol{V})=\emph{Softmax}(\frac{\boldsymbol{Q}\boldsymbol{K}^T}{\sqrt{d}})\boldsymbol{V}
\end{equation}
where the inner product of $\boldsymbol{Q}$ with all $\boldsymbol{K}$ is calculated, we divide each by the scaling factor $\sqrt{\emph{d}}$, and apply a Softmax function to obtain the adaptive attention score whose value depends on the characteristics of the input data, as the weights on $\boldsymbol{V}$. The scaling factor $\sqrt{\emph{d}}$ is to avoid that the gradient of the Softmax function becomes too small when the absolute value of the input matrix is too large to affect the gradient descent. 

Then, as for the multi-head attention, it is the extension of self-attention and we applied it to explore the different types of relevance in different feature spaces of HSI unmixing process. In the HSI unmixing process with $p$ endmembers, rather than using $d_{model}$-dimensional keys, values, and queries to perform a single self-attention mechanism on the HSI tokens, we use different, learned linear projections to project the queries, keys, and values $p$ times to the $d_k$, $d_q$, and $d_v$ dimensions, respectively, to obtain $p$ sub-self-attention heads. For each sub-self-attention head containing the corresponding projected queries, keys, and values, we then execute the self-attention mechanism in parallel to produce the output values of the $d_v$ dimension. As shown in Figure \ref{Fig. 5}(b), the output values of the sub-self-attention heads are concatenated and eventually downscaled again by linear projection so that the output dimension is the same as the initial input dimension. This multi-head attention integrating into the HSI Patch Transformer blocks is shown in Fig. \ref{Fig. 4}(b), and it makes the models jointly focus on the information abundance from different representation subspaces, i.e., from different endmember sub-spaces at different positions. The mathematical expression for multi-head attention is shown as:
\begin{equation}
    \begin{split}
    Head_i=\emph{Self-attention}(\boldsymbol{Q}\boldsymbol{W}_{i}^{\boldsymbol{Q}},\boldsymbol{K}\boldsymbol{W}_{i}^{\boldsymbol{K}},\boldsymbol{V}\boldsymbol{W}_{i}^{\boldsymbol{V}})\\
    MultiHead(\boldsymbol{Q},\boldsymbol{K},\boldsymbol{V})=Concat(head_1,\cdots,head_p)\boldsymbol{W}^O
    \end{split}
\end{equation}
where $\boldsymbol{W}_{i}^{\boldsymbol{Q}}\in\mathbb{R}^{d_{model}\times{d_k}}$, $\boldsymbol{W}_{i}^{\boldsymbol{K}}\in\mathbb{R}^{d_{model}\times{d_k}}$, and $\boldsymbol{W}_{i}^{\boldsymbol{V}}\in\mathbb{R}^{d_{model}\times{d_v}}$ are projections matrices, and $\boldsymbol{W}^{o}\in\mathbb{R}^{pd_{v}\times{d_{model}}}$ is the final linear projection dimensionality reduction matrix. For each of the sub-self-attention head, we adopts $d_k = d_v = d_{model/p} = 64$ so that the total computational cost of muti-head self-attention is analogous to the full dimensional single head self-attention. 

HSI multi-head attention obtains the spatial geometric features of the HSI patch by directly computing the relationship between any two HSI tokens in the HSI patch domain. In addition, the key to HSI Patch Transformer blocks lies in the multi-head attention, where the HSI token is fully connected to be able to model the dependencies between pixels of an HSI patch as shown in Fig. \ref{Fig. 4}. The basic role of the inner product is to determine the relative importance of a single HSI token with respect to all other HSI tokens in the patch domain. Keys, queries, and values simulate a content-based retrieval process using multi-head attention. And the multi-head attention mechanism of HSI Patch Transformer blocks performs two-by-two interactions between HSI tokens, fully utilizing the spectral channel information of a single hyperspectral pixel.

\subsection{Superpixel Segmentation and Random Split for The Synthesis of Hyperspectral Data with Spatial Structure}
Most existing synthetic data generation methods tend to first generate individual abundance vectors that are completely random or obey a certain distribution \cite{ref51, ref52}, and then multiply them with the endmember matrix to obtain the final hybrid hyperspectral pixels. The abundance vectors of similar hybrid hyperspectral pixels with such synthetic method lack structural similarity, which means that the synthetic hyperspectral dataset lacks spatial structure information and reality.

Therefore, we propose a data synthesis method based on superpixel segmentation and random split. This method enables the synthesized hyperspectral data to have a certain spatial structure. Firstly, we unmix a hyperspectral dataset using the traditional method and obtain a rough abundance maps solution. Then, we perform the superpixel segmentation and random segmentation method on the approximate abundance maps solution to get the new abundance maps. With these new abundance maps, we further synthesize the hyperspectral data.

\noindent {\bf{Superpixel segmentation.}}
In order to make the generated abundance maps have a spatial domain structure, we perform simple linear iterative clustering (SLIC) \cite{ref54} on the approximate abundance solution maps, obtaining the superpixel segmentation result. The first step in the clustering process is to initialize the cluster centers, for a single abundance map $\boldsymbol{A}=[a_1,a_2,\cdots,a_N]\in\mathbb{R}^N$ possessing $N=r\times{c}$ abundance points, in which \emph{K} initial cluster centers $C_i=[a_i,r_i,c_i]^T$ are sampled on a regular grid spaced \emph{S} abundance points apart. The grid interval is $S=\sqrt{\frac{N}{K}}$ to produce initial superpixels of equal size. Move the centers to the location with the smallest gradient in the corresponding $3\times3$ neighborhood to avoid the centers falling on the noise point or boundary. Next, in the assignment step, search the nearest cluster center within $2S\times{2S}$ and assign each abundance point to a certain cluster until \emph{K} superpixels are obtained. To determine the nearest cluster center for abundance point $a_i$, we define the abundance distance measure $D_a$ as
\begin{equation}
\label{deqn_x1}
    \begin{split}
    d_c&=\sqrt{(a_i-a_j)^2}\\
    d_s&=\sqrt{(r_i-r_j)^2+(c_i-c_j)^2}\\
    D_a&=\sqrt{(\frac{d_c}{q})^2+(\frac{d_S}{S})^2}
    \end{split}
\end{equation}
where $d_c$ is value distance of the abundance, $d_s$ represents the space distance, $q$ and $S$ are two constants representing the maximum abundance value distance and the maximum spatial distance within the cluster respectively. By defining $D_a$ as (10), the abundance value similarity and spatial proximity are both considered. With each abundance point attached to the nearest cluster center, an update step adjusts the cluster center to the average vector of all pixels belonging to that cluster. The assignment and update steps repeat 10 times in alternation empirically. Then, the superpixel segmentation result of the abundance map is obtained.

\begin{figure}[!t]
\centering
\includegraphics[width=2.5in]{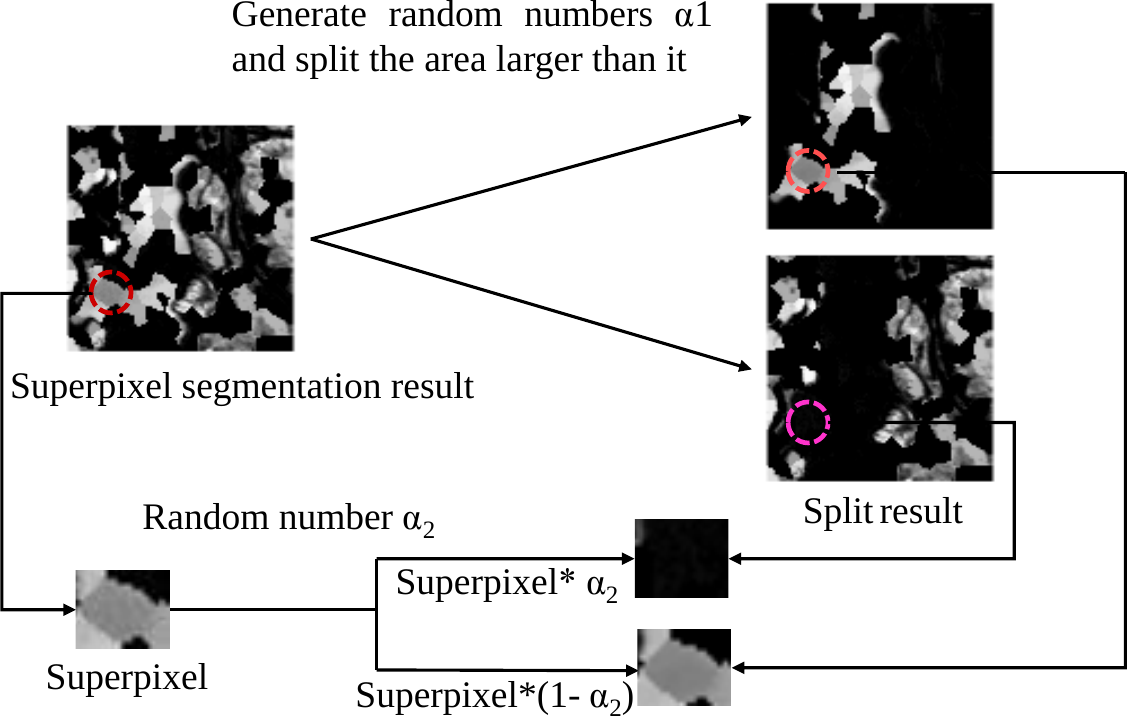}
\caption{Superpixel segmentation and random split for synthetic data.}
\label{Fig. 6}
\end{figure}

\noindent{\bf{Random split.}} To produce new abundance maps, we further randomly split the superpixel segmentation result. It is worth noting that in order to preserve the pure signature structures of endmembers, we avoid splitting the abundance point positions with an abundance value of 1. The split process is depicted in Fig. \ref{Fig. 6}. First, we generate two random variables $\alpha_1$ and $\alpha_2$ between 0 and 1, and define a threshold of 0.5. Assume that the original abundance superpixel image block is $\boldsymbol{P}( p_{i, j}=1, \forall i \in\{1, \ldots, n\}, \forall j \in\{1, \ldots, p\})$, where $n$ is number of pixels contained in a superpixel and $p$ is the number of endmembers. And the two superpixel image blocks after splitting are $\boldsymbol{P_1}$ and $\boldsymbol{P_2}$. The splitting relationship can be denoted as
\begin{align}
{\boldsymbol{P}_1} &=
\begin{cases}
(1-\alpha_2)\boldsymbol{P},&{\alpha_1\ge{0.5} \text { and } p_{i, j} \neq 1, \forall i \in\{1, \ldots, n\}, \forall j \in\{1, \ldots, p\}}\\
{\boldsymbol{P},}&{\alpha_1<0.5 \text { or } p_{i, j}=1, \forall i \in\{1, \ldots, n\}, \forall j \in\{1, \ldots, p\}}
\end{cases}\\
{\boldsymbol{P}_2} &=
\begin{cases}
\alpha_2\boldsymbol{P},&{\alpha_1\ge{0.5} \text { and } p_{i, j} \neq 1, \forall i \in\{1, \ldots, n\}, \forall j \in\{1, \ldots, p\}}\\
{0,}&{\alpha_1<0.5 \text { or } p_{i, j}=1, \forall i \in\{1, \ldots, n\}, \forall j \in\{1, \ldots, p\}}
\end{cases}
.
\end{align}

After all the superpixel blocks are split, the number of abundance maps will be doubled. We then randomly select endmembers from the United States Geological Survey (USGS) \cite{ref42}, The final synthetic Hyperspectral data is obtained by mixing the generated abundance and endmembers and adding Gaussian noise. The mixed model adopted in this paper is the GBM model \cite{ref59}, which is an improved model of the nonlinear model FM model \cite{ref58}. GBM model describes the secondary scattering between objects but ignores the interaction of more than three times scattering, simplifying the model and ensuring the authenticity of true nonlinear hybrid mode. Its mathematical expression is as follows:
\begin{equation}
	\boldsymbol{x}=\boldsymbol{M} \boldsymbol{a}+\sum_{i=1}^{p-1} \sum_{j=i+1}^{p} \gamma_{i . j} \boldsymbol{a}_{i} \boldsymbol{a}_{j} \boldsymbol{m}_{i} \odot \boldsymbol{m}_{j}+\boldsymbol{n}
\end{equation}
where $\gamma_{i, j}$ is the coefficient that controls the interaction between the $\emph{i}th$ and $\emph{j}th$ endmember, $\boldsymbol{n}$ is the noise matrix, $\odot$ represents Hadamard product operation:
\begin{equation}
	\boldsymbol{m}_{i} \odot \boldsymbol{m}_{j}=\left(\begin{array}{lll}
		m_{1, i} & \ldots & m_{L, i}
	\end{array}\right)^{T} \odot\left(\begin{array}{lll}
		m_{1, j} & \ldots & m_{L, j}
	\end{array}\right)^{T}=\left(\begin{array}{lll}
		m_{1, i} m_{1, j} & \ldots & m_{L, i} m_{L, j}
	\end{array}\right)^{T}.
\end{equation}
The constraint conditions of GBM in (13) are as follows:
\begin{equation}
	\begin{gathered}
		a_{k} \geq 0, \forall k \in\{1, \ldots, p\} \\
		\sum_{k=1}^{p} a_{k}=1 \\
		0 \leq \gamma_{i, j} \leq 1, \forall i \in\{1, \ldots, p\}, \forall j \in\{i+1, \ldots, p\}.
	\end{gathered}
\end{equation}
An important property of GBM is that GBM is simplified into LMM model when $\gamma_{i, j}=0$ , into FM model when $\gamma_{i, j}=1$. And when $\gamma_{i, j}$ falls between 0 and 1, the GBM contains both LMM linear mixture and an FM nonlinear mixture. We determine the hybrid mode of the mixed pixels in the synthetic hyperspectral data by adjusting the nonlinear coefficient vector $\gamma=\left[\gamma_{1,2}, \ldots, \gamma_{p-1, P}\right]^{T}$, consequently obtaining more complex synthetic hyperspectral data with more physical meaning to verify the effectiveness of the proposed method.

In the above hyperspectral data synthetic method, the superpixel segmentation can ensure that the generated abundance has a spatial neighborhood structure, which in turn can ensure that the generated hyperspectral data has a certain similarity between the adjacent hyperspectral pixels. And the setting of random variables and thresholds also ensures certain randomness, which produces new abundance maps and synthesizes a new hyperspectral dataset.

\section{Results}

In this section, we implement the proposed unmixing method and carry out the experimental analysis. Synthetic data at different noise levels and real HSI are used to evaluate the effectiveness. To fully demonstrate the performance of our method, we compare it with several typical state-of-the-art demixing methods. In addition, the overall network structure and number of layers of our method are the same in all experiments.

We use four evaluation indicators to assess the accuracy of the experimental results in this paper: the root mean square error (RMSE), the abundance overall root mean square error (aRMSE) \cite{ref32}, root mean square abundance angle distance (rmsAAD) \cite{ref44}, and average spectral angle mapper (aSAM)\cite{ref32}. In the presence of a true abundance label, RMSE measures the numerical similarity of the true abundance $a_j$ of the \emph{j}th endmember to the generated one $\hat{a}_j$ and aRMSE is set to measure the overall mean square error of all HSI pixels, whereas rmsAAD measures dimensional similarity of the \emph{i}th HSI pixel true abundance vector $\boldsymbol{a}_i$ and the generated abundance vector $\hat{\boldsymbol{a}}_i$. These three measurements evaluate the generated abundances from different angles, which makes the evaluation more comprehensive. While the abundance label is unknown, aSAM is adopted to measure the similarity between the original pixel $x_ {i}$ and the reconstructed pixel $\hat{x}_{i}$ to evaluate the unmixing performance. The four evaluation indicators are defined as follows: 
\begin{align}
R M S E_{a_{j}}&=\left(\frac{1}{N}\sum_{i=1}^{N}\left(a_{j,i}-\hat{a}_{j,i}\right)^{2}\right)^{\frac{1}{2}} \\
a R M S E&=\frac{1}{N} \sum_{i=1}^{N}\left(\frac{1}{p}\sum_{j=1}^{p}\left(a_{i,j}-\hat{a}_{i,j}\right)^{2}\right)^{\frac{1}{2}} \\
r m s A A D&=\left(\frac{1}{N}\sum_{i=1}^{N}\left(\arccos \left(\frac{\boldsymbol{a}_{i}^{T} \boldsymbol{a}_{i}}{\left\|\boldsymbol{a}_{i}\right\|\left\|\boldsymbol{a}_{i}^{T}\right\|}\right)\right)^{2}\right)^{\frac{1}{2}}
\end{align}
\begin{equation}
	a S A M=\frac{1}{N} \sum_{i=1}^{N} \arccos \left(\frac{\mathbf{x}_{i}^{T} \hat{\mathbf{x}}_{i}}{\left\|\mathbf{x}_{i}\right\|\left\|\hat{\mathbf{x}}_{i}\right\|}\right).
\end{equation}

The following typical and state-of-the-art algorithms are compared:
\begin{itemize}
\item[1)]
\noindent{The sparse unmixing via variable splitting augmented lagrangian algorithm (SUnSAL):} SUnSAL based on alternating direction multiplier method (ADMM) \cite{ref46} performs unmixing by splitting and augmenting Lagrangian, which searches for the sparsest linear endmember combinations of hyperspectral mixed pixels from a library of known complete endmember spectra. It is an extremely classical method for linear unmixing.
\item[2)]
\noindent{The fully constrained least squares spectral unmixing algorithm (FCLS) \cite{ref66}:} FCLS is a widely used linear unmixing method of figuring out least square problem. It requires two constraints imposed on the linear mixture model applied in the unmixing to estimate the abundance, which are the abundance sum-to-one and the abundance nonnegativity constraints. 
\item[3)]
\noindent{The neural network (NNet) \cite{ref62}:} NNet is a supervised abundance estimation method for nonlinear hyperspectral unmixing, which aims to learn the mapping between pixels spectra and the fractional abundance.
\item[4)]
\noindent{Hyperspectral unmixing using deep image prior (UnDIP) \cite{ref100}:} UnDIP is a deep learning method for the linear unmixing problem, which estimates the abundances using a deep image prior. The main motivation for this work is to improve abundance estimation and enhance robustness to noise.
\item[5)]
\noindent{The deep convolutional autoencoder for hyperspectral unmixing (CAE) \cite{ref63}:} CAE is an end-to-end nonlinear hyperspectral unmixing method based on the convolutional neural network, in which CNN is adopted to utilize spatial-spectral information. This method uses a CNN architecture that consists of two stages: the first stage extracts features and the second stage performs the mapping from the extracted features to obtain the abundance.
\item[6)]
\noindent{The spatial-spectral conditional generative adversarial network for hyperspectral unmixing (scGAN) \cite{ref47}:} scGAN is our previous work that has been published, which is a nonlinear unmixing method. Adversarial strategy and convolution kernel are incorporated into the model to promote the unmixing performance. CNN is adopted to extract features of the spatial-spectral information in the unmixing process.
\end{itemize}

\subsection{Spatial Window Size Analysis of HSI Patch Input}
\begin{figure}[!h]
	\centering
	\includegraphics[height=4.2cm ,width=14cm]{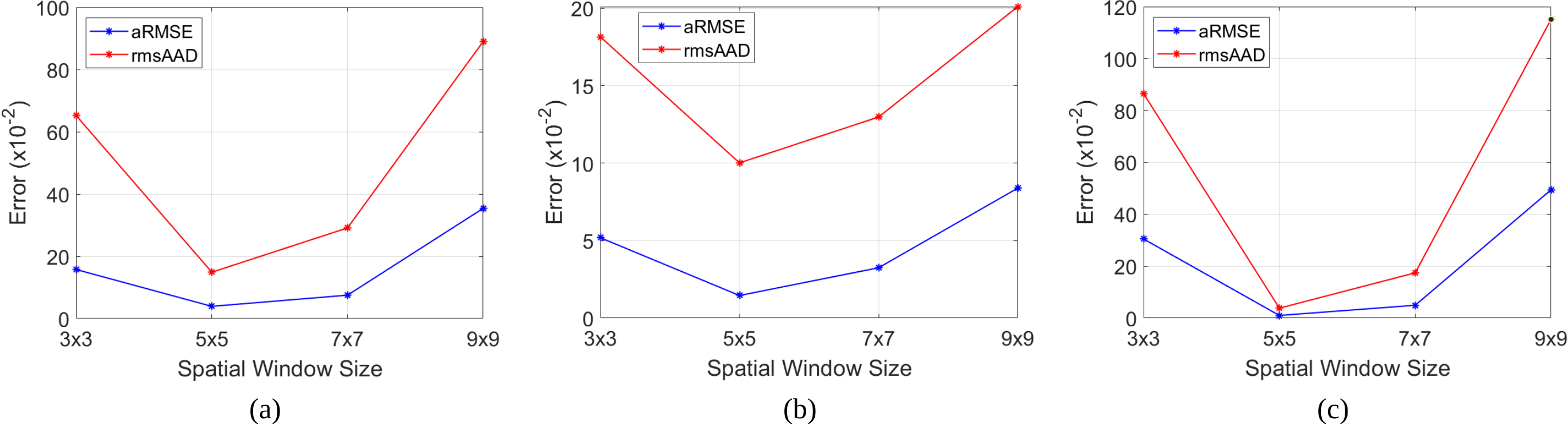}\\
	\caption{ The evaluation ofthe proposed HyperGAN with the different spatial window sizes: (a) Synthetic dataset (SNR = 20dB), (b) Urban dataset, (c) Samson dataset.}
	\label{Fig. 7}
\end{figure}

To locate the optimum spatial window size (s × s), i.e. the HSI patch size, the proposed HyperGAN is investigated with different input HSI patch sizes: 3 × 3, 5 × 5, 7 × 7, 9 × 9. Fig. \ref{Fig. 7} demonstrates the results of the aRMSE and AAD evaluation for the synthetic dataset (SNR = 20 dB), Urban, and Samson adopted the different spatial window sizes. As we can see, spatial window of different sizes will have a significant impact on the unmixing effect. On all datasets, the best unmixing performance is obtained for a size of 5 × 5, thus we chose a spatial window size of 5 × 5. In general, using spatial information from adjacent pixels improves performance in generating central pixel abundance, but increasing the spatial HSI patch size causes a detrimental impact. That is, excessive spatial information of neighborhood pixels may disturb the abundance estimation of central pixels, especially for the pixels in the edges. Also, large spatial window size will bring more network parameters and calculation burden. 
\subsection{Synthetic Dataset Experiments}
\begin{figure}[htb]
	\centering
	\includegraphics[height=4.25cm ,width=13cm]{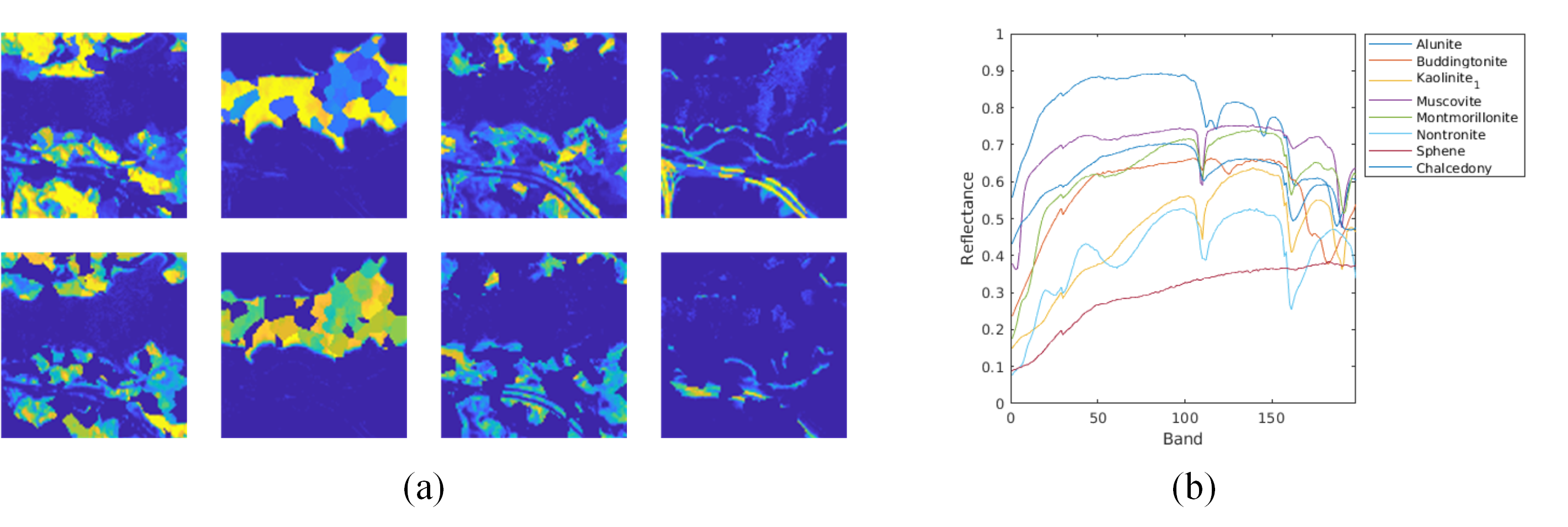}\\
	\caption{Synthetic dataset: (a) Generated abundance maps based on SLIC and random splitting, (b) Endmember graph for synthetic dataset.}
	\label{Fig. 8}
\end{figure}

The synthetic data based on superpixel segmentation and random split method contain eight endmembers randomly chosen from the USGS as shown in Fig. \ref{Fig. 8}(b), which are classified as alunite, buddingtonite, kaolinite1, muscovite, montmorillonite, nontronite, sphene, and chalcedony. Each endmember contains 198 spectral channels. The distribution of generated abundance is shown in Fig. \ref{Fig. 8}(a). In this experiment, the generated abundance and endmember are mixed with the GBM model to obtain the mixed pixels, where the bilinear coefficient matrix $\gamma$ is the matrix with 0.2 elements. Thus there are both linear and nonlinear mixing modes of mixed pixels in synthetic data sets. The synthetic dataset contains 10000 pixels, and 10000 HSI patches for training can be obtained. In this paper, only 4000 HSI patches are randomly selected as labeled data, accounting for 40\% of the total data, and the remaining 60\% of the data are sent to the network as unlabeled data for testing. In the implementation of the proposed method, the number of heads in the multiple attention module h is set to 8, which is equal to the number of endmembers of the synthetic dataset, and each head is anticipated to attend to the abundance of different endmember subspaces. The correction coefficient $\lambda_{cor}$ is set to 10 to guide the abundance generating process of the generator network. The parameters of the Adam optimizer beta1 and beta2 are set at 0.7 and 0.999, respectively. The learning rate is set at 0.0002, which produces the best results. The batch size is tuned to 32, and the spatial window size (S×S) is set (5 × 5).

\begin{figure}[htb]
\includegraphics[width=13cm]{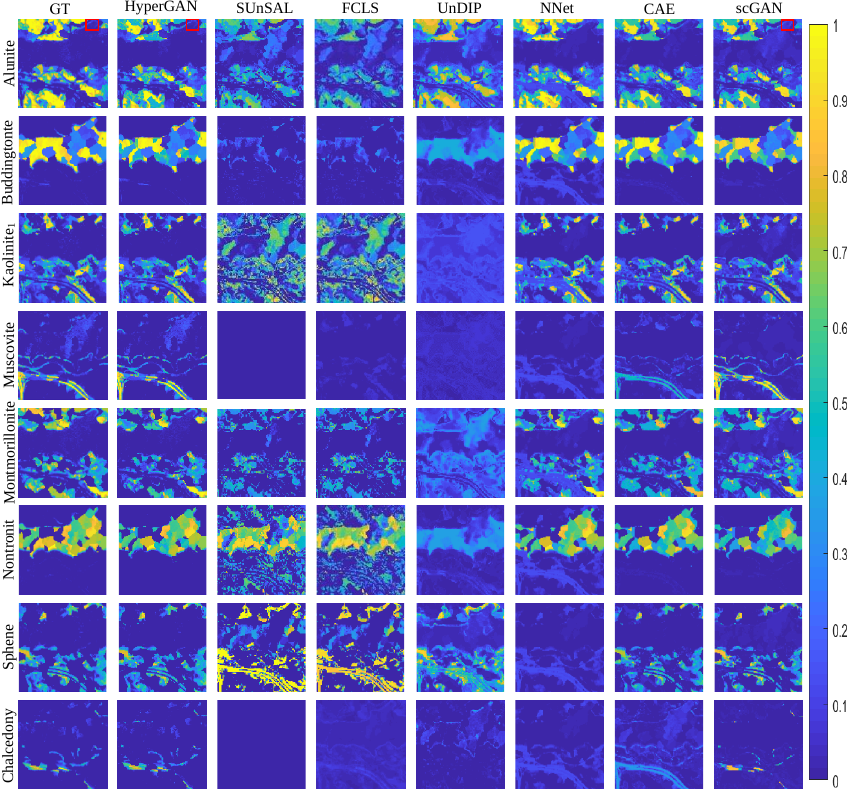}
\caption{Estimated abundance of HyperGAN and comparison algorithms on synthetic dataset (SNR=20dB).\label{Fig. 9}}
\end{figure}   
\unskip

\begin{figure}[htb]
\centering
\includegraphics{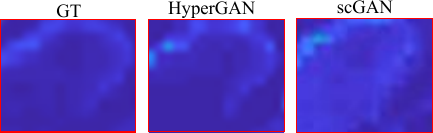}
\caption{Comparison of details in red box of the abundance maps in the first row of the second column and the fifth column from Fig. 9.}
\label{Fig. 10}
\end{figure}

\begin{table}[htbp] 
\caption{Unmixing Evaluation on Sythetic Dataset with Gaussian Noise From SNR = 10-30 dB.\label{Table1}}
	\begin{adjustwidth}{-\extralength}{0cm}
		\newcolumntype{C}{>{\centering\arraybackslash}X}
		\begin{tabularx}{\fulllength}{CCCCCCCC}
			\toprule
			SNR=10dB&HyperGAN&SUnSAL&FCLS&UnDIP&NNet&CAE&scGAN\\
			\midrule
			Alunite&$9.67e-2^\text{a}$&2.74e-1&2.19e-1&2.37e-1&4.82e-2&1.07e-1&1.42e-1\textbf{}\\
			\midrule
			Buddingtonite&1.20e-1&3.29e-1&3.58e-1&4.11e-1&1.07e-1&\bf{9.29e-2}&1.12e-1\\
			\midrule
			Kaolinite1&\bf{3.37e-2}&3.10e-1&3.46e-1&2.07e-1&4.58e-2&5.83e-2&9.88e-2\\
			\midrule
			Muscovite&bf{4.22e-2}&1.54e-1&1.87e-1&1.88e-1&1.62e-1&7.34e-2&1.04e-1\\
			\midrule
			Montmorillonite&\bf{1.24e-1}&2.70e-1&2.81e-1&3.07e-1&3.01e-1&1.49e-1&1.57e-1\\
			\midrule
			Nontronite&\bf{7.25e-2}&4.14e-1&3.20e-1&3.99e-1&1.03e-1&1.02e-1&1.03e-1\\
			\midrule
			Sphene&\bf{3.17e-2}&4.19e-1&3.52e-1&4.07e-1&5.19e-2&7.11e-2&1.28e-1\\
			\midrule
			Chalcedony&\bf{3.73e-2}&1.43e-1&1.05e-1&1.28e-1&9.58e-2&9.76e-2&1.02e-1\\
			\midrule
			aRMSE&\bf{4.35e-2}&2.73e-1&2.61e-1&2.78e-1&9.95e-1&7.72e-2&9.58e-2\\
			\midrule
			rmsAAD&\bf{1.70e-1}&1.14&1.05&1.23&5.17e-1&2.88e-1&4.02e-1\\
	        \bottomrule
		\end{tabularx}
		
	    \begin{tabularx}{\fulllength}{CCCCCCCC}
			\toprule
			SNR=20dB&HyperGAN&SUnSAL&FCLS&UnDIP&NNet&CAE&scGAN\\
			\midrule
			Alunite&1.25e-1&2.12e-1&2.11e-1&3.57e-1&\bf{7.79e-2}&1.22e-1&1.23e-1\\
			\midrule
			Buddingtonite&1.18e-1&3.46e-1&3.44e-1&3.39e-1&1.17e-1&\bf{9.57e-2}&1.31e-1\\
			\midrule
			Kaolinite1&\bf{1.98e-2}&3.34e-1&3.39e-1&2.23e-1&4.60e-2&6.84e-2&5.86e-2\\
			\midrule
			Muscovite&\bf{2.33e-2}&1.87e-1&1.88e-1&1.76e-1&1.60e-1&1.19e-1&6.13e-2\\
			\midrule
			Montmorillonite&1.15e-1&2.51e-1&2.54e-1&2.79e-1&1.62e-1&\bf{1.04e-1}&1.30e-1\\
			\midrule
			Nontronite&\bf{6.79e-2}&3.13e-1&3.08e-1&3.14e-1&9.69e-2&9.51e-2&7.54e-2\\
			\midrule
			Sphene&\bf{1.79e-2}&3.73e-1&3.39e-1&1.98e-1&1.86e-1&5.24e-2&6.59e-2\\
			\midrule
			Chalcedony&\bf{1.80e-2}&1.05e-1&1.06e-1&1.30e-1&9.42e-2&9.85e-2&9.26e-2\\
			\midrule
			aRMSE&\bf{3.94e-2}&2.53e-1&2.50e-1&2.52e-1&9.09e-2&7.38e-2&7.77e-2\\
			\midrule
			rmsAAD&\bf{1.49e-1}&1.03&1.02&1.05&4.84e-1&2.85e-1&2.86e-1\\
			\bottomrule
		\end{tabularx}
		
		\begin{tabularx}{\fulllength}{CCCCCCCC}
			\toprule
			SNR=30dB&HyperGAN&SUnSAL&FCLS&UnDIP&NNet&CAE&scGAN\\
			\midrule
			Alunite&1.32e-1&3.20e-1&2.10e-1&3.62e-1&8.70e-2&\bf{1.20e-1}&1.34e-1\\
			\midrule
			Buddingtonite&1.17e-1&3.06e-1&3.43e-1&3.41e-1&\bf{6.09e-2}&9.07e-2&9.45e-2\\
			\midrule
			Kaolinite1&\bf{1.33e-2}&2.16e-1&3.38e-1&2.33e-1&4.23e-2&5.30e-2&5.28e-2\\
			\midrule
			Muscovite&\bf{1.83e-2}&1.33e-1&1.87e-1&1.83e-1&1.60e-1&6.94e-2&6.00e-2\\
			\midrule
			Montmorillonite&1.18e-1&3.07e-1&2.56e-1&2.78e-1&1.40e-1&1.15e-1&\bf{1.08e-1}\\
			\midrule
			Nontronite&\bf{6.68e-2}&3.01e-1&3.07e-1&3.12e-1&1.68e-1&9.92e-2&1.18e-1\\
			\midrule
			Sphene&\bf{1.49e-2}&2.21e-1&3.40e-1&1.98e-1&1.89e-1&5.90e-2&5.54e-2\\
			\midrule
			Chalcedony&\bf{1.37e-2}&1.31e-1&1.05e-1&1.38e-1&8.57e-2&9.92e-2&5.26e-2\\
			\midrule
			aRMSE&\bf{3.82e-2}&2.10e-1&2.49e-1&2.57e-1&9.20e-2&7.03e-2&7.04e-2\\
			\midrule
			rmsAAD&\bf{1.53e-1}&9.71e-1&1.02&1.07&4.86e-1&2.49e-1&2.39e-1\\
			\bottomrule
		\end{tabularx}
	\end{adjustwidth}
	\noindent{\footnotesize{\textsuperscript{a} In this article denotes RMSE for endmember abundance errors.}}
\end{table}
\emph{1) Comparison with the State-of-the-Art Unmixing Algorithms:} The comparison of the unmixing abundance results for the synthetic dataset (SNR=20dB) on different algorithms is shown in Fig. \ref{Fig. 9}. Observing the ground truth (the first column of Fig. \ref{Fig. 9}), the abundance maps generated by superpixel segmentation and random split have obvious spatial structure characteristics. As for the unmixing result, SUnSAL and CAE both show excellent performance at unmixing. However, the proposed HyperGAN excels in detail, especially the edges of the abundance map look relatively clean and accurate, close to the real maps. Comparing the unmixing results between HyperGAN and scGAN, it can be seen that the latter model considering spatial information only by convolution kernel is more likely to lead to excessive homogeneity of local regions, such as the Fig. \ref{Fig. 10} show that the comparison of details in red box of the abundance maps in the first
row of the second column and the fifth column from Fig. \ref{Fig. 9} , which shows the effectiveness of our model.

\emph{2) Analysis of Robustness to Noise:} We performed experiments to compare and investigate the robustness of the proposed method. In this experiment, the synthetic dataset is contaminated with varying degrees of white Gaussian noise, where the SNR varied from 10 dB, 20 dB, and 30 dB. A quantitative analysis of these algorithms is presented in Table \ref{Table1}. Overall, all the algorithms show a decrease in the accuracy of the unmixing as the noise level increases. We can note that the proposed method shows better robustness in the three metrics of RMSE, aRMSE, and rmsAAD. However, for the autoencoder-based method NNet, which unmixes for individual hyperspectral pixels, the unmixing effect fluctuates widely from 20dB to 10dB in noise level and is more affected by noise. For CAE, which uses convolutional networks to unmix using joint spatial spectral information, the unmixing performance of CAE is favorable when the noise level of the synthetic dataset is SNR = 30dB. In particular, the improvement is significant when compared to higher noise levels (20dB). For the traditional algorithms MVCNMF and SUnSAL, VCA initialisation can be influenced by noise levels, which can affect the unmixing effect. For the result of our previous work scGAN at different noise levels, the unmixing effect is satisfactory with little fluctuation. Nevertheless, the proposed method uses HSI Patch Transformer block instead of convolution kernel to exploit the spatial information of HSI a more fine-grained way, which leads to that the proposed method significantly outperforms the scGAN in each measurement index. In addition, the proposed HyperGAN also surpasses other methods and yields the optimal or equivalent results in most cases, which demonstrates the effectiveness and robustness of our method.

\emph{3) Impact of the HSI Patch Transformer blocks:} To illustrate that multi-head attention of HSI Patch Transformer block captures HSI patch space features more finely, in this experiment, we randomly selected an HSI patch from a superpixel dataset consisting of eight endmembers and analyzed the process of the estimated abundance of the central pixel. Herein, the multi-head attention is accordingly set to eight heads.

\begin{figure*}[!t]
\centering
\includegraphics{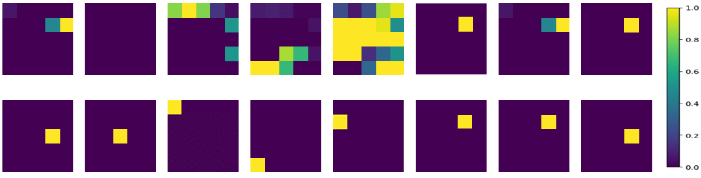}
\caption{(The first row) The true abundance of an HSI patch. (The second row) The central pixel attention score of the HSI patch computed by the multi-head layer ahead of the estimated abundance layer.}
\label{Fig. 11}
\end{figure*}

First, we input a randomly selected HSI patch into our trained proposed model, expecting to get the corresponding abundance output of the central pixel. In the meantime, the extracted adaptive attention score graph of the central pixel is demonstrated in the second row of Fig. \ref{Fig. 11}. And the first row is the true abundance maps of the HSI patch possessing physical characteristics of sparsity. Note that as for the HSI patch, we aim to take advantage of the collaborative spatial-spectral information to help for abundance estimation improvement of the central pixel, therefore focusing on the relationship between the central HSI pixel and the domain pixels. As shown in Fig. \ref{Fig. 11}, in the different endmember subspaces, the regions whose abundance is not zero are assigned high fraction values by the multi-head attention. The remaining regions of low attention score in that endmember subspace are the so-called edge regions within the corresponding abundance maps, which means these regions are composed of other endmembers. Therefore, it is proved that in the well-trained model, the adaptive attention scores obtained from the interaction of the central pixel and the domain pixels in the multi-head attention layer predict well the most likely locations of the abundance of the different endmembers. While in the second column of Fig. \ref{Fig. 11}, the true abundance value for all positions of this endmember is zero, but the highest attention score value is at the central pixel position of the HSI patch, indicating that the model learns to give aprioroty to the central pixel. Based on the above analysis, we can infer that the multi-head attention module has learned to pay attention to the central pixel and get spatial distribution information of abundance, which is consistent with our expected results.
\subsection{Real Dataset Experiments}
\emph{1) Urban:} The Urban dataset \cite{ref48} is an HSI dataset used extensively in HSU research. It contains 307 x 307 pixels and 210 spectral bands ranging from 400 nm to 2500 nm. With bad bands (1-4, 78, 87, 101-111, 136-153, 198-210) removed, there are 162 bands left in the Urban dataset. The urabn dataset has four endmembers: asphalt, grass, tree, and roof. The urban dataset contains 94249 pixels, only 10\% of the total data is randomly selected as labeled data in our training, and the remaining 90\% of the data are sent to the network as unlabeled data for testing. Regarding the parameter setting, the number of heads in the multiple attention module h is set to 4. The correction coefficient $\lambda_{cor}$ is set to 5. The beta1, beta2 and learning rate of the Adam optimizer are set to 0.5, 0.999 and 0.0002 respectively. The batch size is tuned to 32, and the spatial window size is set 5 × 5.

\begin{figure}[htb]
\includegraphics[width=12.5cm]{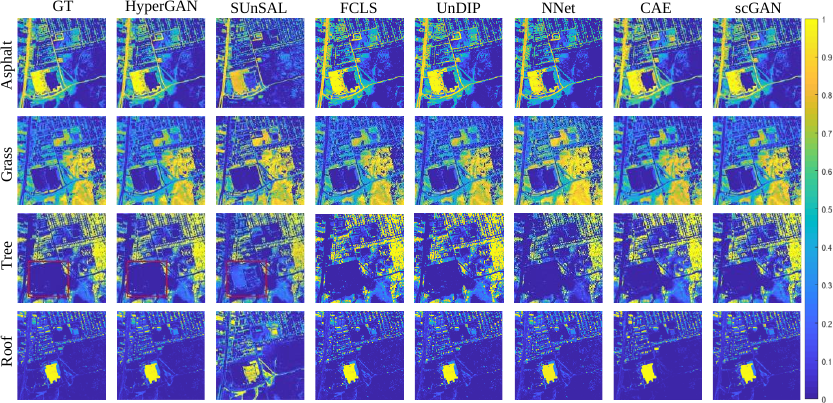}
\caption{Estimated abundance of HyperGAN and comparison algorithms on Urban dataset. \label{Fig. 12}}
\end{figure}

Fig. \ref{Fig. 12} shows the results of the abundance maps estimated by each algorithm. As shown in the red box in the third row of the Fig. \ref{Fig. 12}, compared to other methods, the HyperGAN results guarantee both the sparse characteristics of the abundance of the region and maintain unmixing accuracy. Specifically, SUnSAL shows deficiencies in this case, with fuzzy residues in the sparse region. Note that SUnSAL is a sparse regression method based on a linear mixed model, which only considers $L_1$ sparsity without considering spatial information. For other autoencoder-based networks, the proposed HyperGAN approach is also a credible enhancement. In addition, Our approach is noticeably superior to previous work scGAN, because the muti-head attention mechanism further utilizes the rich spectral band interaction information. Table \ref{Table2} describes the quantitative results of the compared methods and demonstrates the validity of our approach.

\unskip

\begin{table}[H] 
\caption{Unmixing Evaluation Comparison on Urban Dataset.\label{Table2}}
\newcolumntype{D}{>{\centering\arraybackslash}X}
\begin{tabularx}{\textwidth}{CCCCCCCC}
\toprule
Material &HyperGAN&SUnSAL&FCLS&UnDIP&NNet&CAE&scGAN\\
\midrule
Asphalt	&3.31e-2&2.44e-1&2.40e-1&\bf{2.62e-2}&6.37e-2&7.94e-2&3.90e-2\\
\midrule
Grass&\bf{3.47e-2}&1.63e-1&1.64e-1	&3.97e-2	&1.24e-1&7.72e-2&4.53e-2\\
\midrule
Tree &\bf{1.89e-2}&1.48e-1	&1.45e-1&2.70e-2&1.09e-1&6.82e-2&2.93e-2\\
\midrule
Roof&1.95e-2 &2.82e-1&2.82e-1	&\bf{1.51e-2}&3.24e-2&7.43e-2&3.10e-2\\
\midrule
aRMSE&\bf{1.45e-2}&1.68e-1	&1.67e-1&2.37e-2&7.49e-2&5.33e-2&2.36e-2\\
\midrule
rmsAAD	&\bf{1.00e-1}&5.93e-1 &5.83e-1	&7.26e-2&2.58e-1&1.91e-1&1.04e-1\\
\bottomrule
\end{tabularx}
\end{table}
\unskip

\emph{2) Samson:} Samson \cite{ref48} is a simple dataset used in the HSU research. It possesses 95×95 pixels and 156 spectral bands covering the wavelengths from 401 nm to 889 nm. This dataset is not degraded by blank or heavily noisy spectral bands, so there are no spectral bands that require removal. The samson dataset has three endmembers: soil, tree, and water. The samson dataset contains 9025 pixels, only 10\% of the total data is randomly selected as labeled data in our training, and the remaining 90\% of the data for testing. The number of heads in the multiple attention module h is set to 3. The correction coefficient $\lambda_{cor}$ is set to 5. The Adam optimizer with a learning rate of 0.0002 is utilized to optimize the problem, and the parameters beta1 and beta2 are 0.5 and 0.999. The batch size is tuned to 32, and the spatial window size is set 5 × 5.

\begin{figure}[htb]
\includegraphics[width=12.5cm]{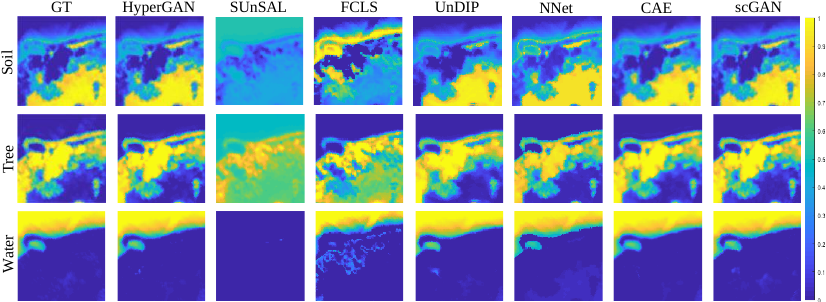}
\caption{Estimated abundance of HyperGAN and comparison algorithms on Urban dataset. \label{Fig. 13}}
\end{figure}

Comparison results of the estimated abundance maps are shown in Fig. \ref{Fig. 13}. We can observe that our HyperGAN method yields comparatively accurate estimates, regardless of whether it is in highly mixed or homogeneous regions, i.e., soil and water. Table \ref{Table3} gives the results of the quantitative estimates for the compared methods the best performance is in bold. Regarding the endmember MSE, aRMSE, and rmsAAD measurement metrics, the unmixing performance of our HyperGAN method stands first of the methods considered in the experiment. 

\unskip

\begin{table}[H] 
\caption{Unmixing Evaluation Comparison on Samson Dataset Umixing.\label{Table3}}
\newcolumntype{C}{>{\centering\arraybackslash}X}
\begin{tabularx}{\textwidth}{CCCCCCCC}
\toprule
Material&	HyperGAN&SUnSAL&FCLS&UnDIP&NNet&CAE&scGAN\\
\midrule
Soil&\bf{2.40e-2}&5.18e-1&5.17e-1&	8.74e-2&	7.63e-2&4.40e-2&2.73e-2\\
\midrule
Tree&	\bf{1.59e-2}1&3.81e-1&	3.33e-1&9.42e-2&	5.60e-2&4.66e-2&2.33e-2\\
\midrule
Water&	\bf{1.13e-2}&3.31e-1&3.27e-1&	4.56e-2&	4.96e-2&2.02e-2&1.83e-2\\
\midrule
aRMSE&\bf{9.87e-3}&3.76e-1&	3.61e-1&4.92e-2&	5.39e-2&2.40e-2&1.45e-2\\
\midrule
rmsAAD&	\bf{3.80e-2}&9.49e-1&9.39e-1&1.74e-1&	1.25e-1&8.51e-2&5.24e-2\\
\bottomrule
\end{tabularx}
\end{table}
\unskip

\emph{3) Cuprite:} 
The dataset was collected by the 1997 Airborne Visible Infrared Imaging Spectrometer (Aviris) and covers the Cuprite region of Nevada, USA.
The raw images have 224 bands ranging from 370nm to 2480nm and a spatial resolution of The spatial resolution is 20m. A sub-image of the Cuprite image with 250 x 190 pixels is used. Of these, 1-2, 104-113, 148-167 bands and 221-224 were removed due to low absorbance and low signal-to-noise ratio. The remaining 188 bands of the dataset are used for experimental analysis. There are no true abundance labels, with true endmember labels\cite{ref60}. The number of heads in the multiple attention module h is set to 12. The correction coefficient $\lambda_{cor}$ is set to 10. The beta1, beta2 and the learning rate of Adam optimizer are set at 0.7, 0.999 and 0.0002. The batch size is tuned to 16, and the spatial window size is set 5 × 5.

For the unmixing of Cuprite dataset, we first randomly selected 30\% mixed pixels from this dataset and obtain the corresponding HSI patches. Then we adopt collaborative nonnegative matrix factorization\cite{ref61} (CoNMF) to unmix these mixed pixels, which is a simple traditional unmixing method based on nonnegative matrix. The unmixing result obtained is used as an approximation of the true abundance label to train the proposed HyperGAN and the remaining mixed pixels are unmixed through our trained HyperGAN. As the Cuprite dataset has the endmember label but no abundance label, the aSAM metric is utilized to measure the unmixing performance by comparing the reconstructed pixels and the original pixels. The quantitative unmixing evaluation is shown in the Table \ref{Table4}.

According to Table \ref{Table4}, the proposed HyperGAN method achieves the optimal unmixing result, and the aSAM value is 0.1513. Compared with the scGAN method, the results of the proposed algorithm improved by 6.9\%. It can be concluded that the Patch Transformer module is superior to the convolutional neural network module in its ability to exploit the spatial-spectral joint information in the unmixng process.
\begin{table}[H] 
\caption{Unmixing Evaluation Comparison on Cuprite Dataset.\label{Table4}}
\newcolumntype{C}{>{\centering\arraybackslash}X}
\begin{tabularx}{\textwidth}{CCCCCCCC}
\toprule
{}&HyperGAN&SUnSAL&FCLS&UnDIP&NNetE&CAE&scGAN\\
\midrule
aSAM&\bf{1.51e-1}&1.75e-1	&1.92e-1	&1.55e-1&1.69e-1&1.85e-1&1.63e-1\\
\bottomrule
\end{tabularx}
\end{table}
\unskip
\subsection{Ablation study} 
To analyze the effectiveness of the discriminator networks of HyperGAN, we further conduct ablation experiments in terms of discarding the key module, i.e., the discriminator networks (denoted as w/o Discriminator). We report the aRMSE and AAD ablation results in Tables \ref{Table5} and \ref{Table6} for the synthetic and real datasets, respectively, for a total of five datasets. Obviously, the discriminator component of the HyperGAN contributes to the improvement to unmixing. As these two tables depict, the value of aRMSE and AAD increase significantly when discarding the discriminator module, especially for the synthetic datasets mixing eight endmembers. And the AAD ablation result on the urban dataset is similar. It may be that the mixing method of the urban dataset is relatively simple, because its unmixing errors are very small, resulting in the difficulty of widening the gap. These results are within our expectation, as discarding the critical discriminator module would result in the model losing its adversarial strategy, which makes it hard for the generator to simulate a suitable unmixing model to generate abundance, especially in complex mixing scenarios. This verifies that the effect of the generator to generate abundance can benefit from the resistance of the discriminator module. 

\begin{table}[H] 
\caption{The aRMSE Ablation Results on the Synthetic Dataset and the Real Dataset.\label{Table5}}
\newcolumntype{C}{>{\centering\arraybackslash}X}
\begin{tabularx}{\textwidth}{CCCCCC}
\toprule
{}&	Synthetic &Synthetic &Synthetic &{}&{}\\
Methods&dataset&dataset&dataset&Urban&Samson\\
{}&/SNR=10dB&/SNR=20dB&/SNR=30dB&{}&{}\\
\midrule
\bf{Ours}&	\bf{4.35e-2}&	\bf{3.94e-2}&	\bf{3.82e-2}&	\bf{1.45e-2}&	\bf{9.87e-3}\\
\midrule
w/o Discriminator&	1.58e-1	&2.76e-1&	9.87e-2	&4.26e-2&	1.02e-2\\
\bottomrule
\end{tabularx}
\end{table}
\unskip

\begin{table}[H] 
\caption{The rmsAAD Ablation Results on the Synthetic Dataset and the Real Dataset.\label{Table6}}
\newcolumntype{C}{>{\centering\arraybackslash}X}
\begin{tabularx}{\textwidth}{CCCCCC}
\toprule
{}&	Synthetic &Synthetic &Synthetic &{}&{}\\
Methods&dataset&dataset&dataset&Urban&Samson\\
{}&/SNR=10dB&/SNR=20dB&/SNR=30dB&{}&{}\\
\midrule
\bf{Ours}&	\bf{1.70e-1}&	\bf{1.49e-1}&	\bf{1.53e-1}&	\bf{1.00e-2}&	\bf{5.23e-2}\\
\midrule
w/o Discriminator&	4.24e-1	&8.28e-1&	3.27e-1	&1.27e-1&	11.51e-2\\
\bottomrule
\end{tabularx}
\end{table}
\unskip

\section{Discussion}
From the above experimental results, the performance of each algorithm is different in different datasets. Overall, the proposed HyperGAN algorithm achieves the best unmixing results in terms of aRMSE, rmsAAD, and aSAM metrics, as well as greater robustness. The main reasons for the superior performance of the HyperGAN algorithm are as follows. Firstly, the modality transformation from hyperspectral image patch to the abundance of corresponding endmembers of the central pixels is based on the generator consisting of the proposed Patch Transformer blocks, which extracts the joint spatial-spectral information to achieve the unmixing, and outperforms the NNet algorithm which only uses the spectral information. The Patch Transformer module also has an advantage over the CNN, as evidenced by the experimental results comparing HyperGAN and scGAN. After the algorithm is trained, the weights of the CNN fused features are fixed, and the features of the central hyperspectral pixels and their neighboring hyperspectral pixels are fused equally for the unmixing of the central hyperspectral pixels. Nevertheless, the structure of hyperspectral image patches is multivariate, and the neighboring hyperspectral pixels of different hyperspectral image patches have different correlations with the corresponding central hyperspectral pixels, and the weights for fused features should be adaptive. The adaptive attention scores are used as fusion feature weights, and the weights of the fused features depend to some extent on the characteristics of the data itself, thus is adaptive. The proposed Patch Transformer uses the collaborative spatial-spectral information in a fine-grained way to generate the abundance of the central hyperspectral pixels, to optimize the unmixing process. Secondly, HyperGAN uses a discriminator composed of fully connected layers to determine whether the structure and distribution of the abundance obtained from the unmixing are consistent with the true abundance. Unlike general deep learning algorithms, the discriminator is equivalent to a complex loss term that adapts to different unmixing scenarios and different hyperspectral datasets, effectively improving the robustness of the algorithm and eliminating the need to manually select prior knowledge to set the loss function. The results of comparison experiments with each algorithm and ablation experiments demonstrate that the discriminator key module plays an important role in the unmixing performance. The generator and discriminator play against each other using an adversarial strategy and evolve to optimize the unmixing performance.
	
The proposed method also has some limitations. As the proposed method is based on the supervised unmixing scenarios, future work will extend the proposed framework to unsupervised unmixing scenarios.

\section{Conclusions}
Most unmixing methods utilize prior knowledge about abundance distribution to solve complex regularization optimization problems. To eliminate the troublesome selection problem of prior knowledge and avoid solving complex regularization optimization problems, this paper proposes a supervised hyperspectral unmixing method that is the conditional generative adversarial network based on the Patch Transformer, which provides a new perspective to solve the unmixing problem: the unmixing process from pixel to abundance the unmixing process can be regarded as a transformation of two modalities. The proposed method mainly consists of two networks: a generator and a discriminator, the former completes the modal conversion from mixed hyperspectral pixel patch to the abundance of corresponding endmembers of the central pixel and the latter is used to distinguish whether the distribution and structure of generated abundance are the same as the true ones. Aiming to adaptively utilize the spatial correlation information of pixels in hyperspectral image patch, the Patch Transformer module of generator is designed, in which the adaptive attention score of the central pixel feature and adjacent pixel feature is calculated to assist in producing efficient representation of features. Our method is experimentally evaluated using synthetic and real hyperspectral data, and the experimental results show the superiority of our method compared with several classical and state-of-the-art methods. As the proposed method is based on the supervised unmixing scenarios, future work will extend the proposed framework to unsupervised unmixing scenarios.

\authorcontributions{All authors made great contributions to the work. Conceptualization, H.Z. and L.W.; methodology, L.W.; software, L.W.; validation, H.Z., and L.W.; formal analysis, L.W.; investigation, L.W.; resources, H.Z.; data curation, L.W. and F. L.; writing---original draft preparation, L.W.; writing---review and editing, H.Z., F.L., H.C., and Y.M.; visualization, L.W.; supervision, X.X.; project administration, H.Z.; funding acquisition, H.Z. All authors have read and agreed to the published version of the manuscript.}

\funding{This research was funded in part by the National Natural Science Foundation of China under Grant 61877066, Aero-Science Fund under Grant 20175181013 and Science and technology plan project of Xi'an under Grant 21RGZN0010.}

\institutionalreview{Not applicable.}

\informedconsent{Not applicable.}

\conflictsofinterest{The authors declare no conflict of interest.} 

\begin{adjustwidth}{-\extralength}{0cm}

\reftitle{References}

\end{adjustwidth}
\end{document}